\documentclass[11pt,twoside]{article}

\usepackage{asp2004}
\usepackage{epsf}
\usepackage{psfig}
\usepackage{lscape}
\usepackage{graphicx}

\begin{document}
\title{Distribution of extinction and star formation in NGC~1569}  
\author{M. Rela\~no, U. Lisenfeld, E. Battaner}
\affil{Dpto. F\'\i sica Te\'orica y del Cosmos, Universidad de
  Granada, Avda. Fuentenueva s/n, 18071, Spain}
\author{J.M. Vilchez}
\affil{Instituto de Astrof\'\i sica de Andaluc\'\i a, CSIC, 
Apartado 3004, 18080, Granada, Spain}    
\author{P. P\'erez-Gonz\'alez}
\affil{Steward Observatory, University of Arizona, 933 North Cherry
  Avenue, Tucson, AZ 85721}

\begin{abstract}
We investigate the distribution of the intrinsic extinction in
NGC~1569 using an extinction map derived from the H$\alpha$/H$\beta$
emission line ratio. We compare the extinction distribution to that of
the dust emission traced by SPITZER IRAC(8~$\mu$m) and MIPS (24~$\mu$m) maps. 
The intrinsic extinction shows spatial variations, from zones with 
negligible extinction to zones with values up to A$\rm_V$=0.8~mag. 
We find an extinction shell and
establish a relation between this shell and the interstellar expanding
structure produced by the stellar winds comig from the Super Star
Cluster (SSCs) A and B in the center of the galaxy. We suggest that the extinction shell
has been produced by the accumulation of dust at the border of the shell.  
Although we find a good spatial correlation between the Balmer
extinction and infrared emission, there is a spatial displacement between the 8~$\mu$m and 
24~$\mu$m maxima and the maximum in Balmer extinction which needs
further investigation. 

\end{abstract}

\section{Introduction}
NGC~1569 is a dwarf starburst galaxy which hosts two Super Star Clusters (SSCs) A and B at
its center. The lack of ionized gas around the SSCs has been ascribed
by several authors to strong stellar winds and supernova explosions
produced by these clusters. Using long-slit echelle H$\alpha$ spectra,
Martin (1998) identified an expanding shell of ionized gas centered
at the position of these SSCs. The distance of the galaxy (2.2~Mpc;
Israel 1988) allows adequate spatial resolution and makes it an ideal object to
study the distribution of the interstellar dust and gas under 
the action of the strong stellar winds and supernova produced by the central SSCs. 

A common way of estimating the extinction caused by the interstellar
dust is the study of the differences between the observed Balmer
decrement and the theoretical expected value. Detailed observations of 
dust emission have been difficult so far but new observations from
SPITZER are presently
producing improvements which will be complemented with the missions
HERSCHEL and PLANCK. 

NCG~1569 is significantly contaminated by
foreground Galactic extinction due to its low Galactic latitude. 
Based on previous studies of the Galactic contribution to
the extinction of NGC~1569 (see Israel 1988; Origlia et al. 2001;
Devost et al. 1997 and Burstein \& Heiles 1984) we assume here a value
of the Galactic extinction of $\rm A_{\scriptsize V}=1.64$~mag. ($\rm
A_{\scriptsize H\alpha}=1.28$).

\section{Extinction map}   

The H$\alpha$ extinction ($\rm A_{\scriptsize H\alpha}$) map was calculated using H$\alpha$ and H$\beta$
images obtained with the WFPC2 on the Hubble Space Telescope and
following the method explained in Caplan \& Deharveng (1986). In order
to study the distribution of the extinction at the global scale of the
galaxy and its relation to the star formation and the kinematic
features of Martin (1998), we have smoothed the original images to a
lower resolution of 6'' ($\sim$65pc in NGC~1569). 

\begin{figure}
   \centering
   \includegraphics[width=9.2cm]{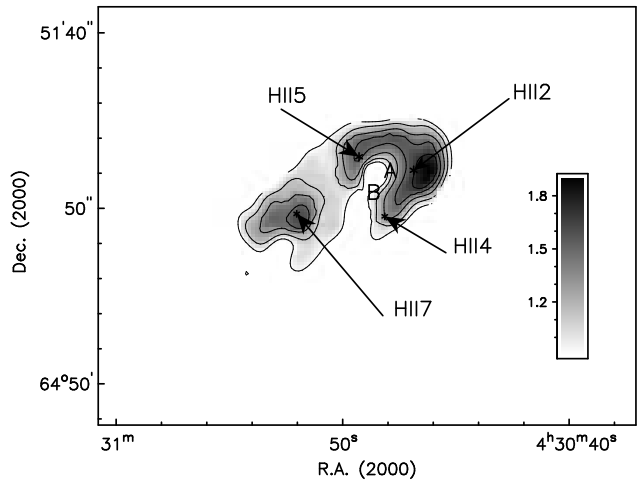}
   \caption{Extinction map (6\arcsec) for the galaxy NGC~1569 obtained
   as a composition of the extinction for the diffuse gas and the
   HII regions. The extinction shown is the sum of intrinsic and
   Galactic extinction. The extinction contours are at $\rm A_{\scriptsize H\alpha}=$0.9, 1.1, 1.3, 1.5,
   1.7 magnitudes and the bar shows the adopted grayscale. 
The extinction values could be overestimated up to a maximum of
   0.05~mag due to the underlying stellar absorption (see Rela\~no et
   al 2006).}
              \label{extHaHb}
    \end{figure}

An extinction map of the galaxy was obtained in two different regimes
characterized by different physical conditions: extinction 
within the HII regions and extinction of the
diffuse ionized gas (DIG).
For a detailed explanation of the procedure see
Rela\~no et al. (2006). In Fig.~\ref{extHaHb} we show the combined
extinction map. The absolute extinction
shows considerable variation over the disk of the galaxy. Once the
Galactic extinction is subtracted, we find values up to 
$\rm A_{\scriptsize H\alpha}=0.6$~mag, corresponding to an extinction
in the V-Band of $\rm A_{\scriptsize V}=0.8$. This result agrees with values
previosly obtained by Kobulnicky \& Skillman (1997).

\subsection{Origin of the extinction distribution}	
The most prominent structure visible in
Fig.~\ref{extHaHb} is the arc formed by the highest values of
extinction (values between 1.5~mag and 1.9~mag), which resembles a shell located
near the position of the HII region 2 of Waller (1991). This {\it
  shell structure} is centered around the position of the Star Cluster A, 
with an average radius of 90~pc, and correlates spatially with the supershell NGC~1569-C catalogued by
Martin (1998). We suggest that the shell extinction structure has
been formed by a cumulative deposit of dust at the boundary
of the supershell. An estimate of the kinematic
age for the supershell is consistent within the age of
the Star Cluster A ($\tau\geq$7~Myr, see Hunter et al. 2000) and an
analysis of the energies involved shows that the stellar winds
coming from SSC A are able to produce the expanding shell and to
sweep a significant amount of dust mass contained in it (Rela\~no et
al. 2006). 

\section{Balmer extinction and distribution of the interstellar gas and dust}

The comparison of the extinction map with the H$\alpha$ emission done
by Rela\~no et al. (2006) shows that the maximum in extinction within the shell
structure does not coincide spatially with the maximum in H$\alpha$
emission, which corresponds to the position of the HII region 2 (see
Fig.~\ref{extHaHb}). 
The H$\alpha$ maximum is closer to the position of SSC
A than the maximum in extinction, which is
displaced by $\sim$6'' to the west. We interpret the displacement 
as a consequence of dust accumulating 
in the outer region of the ionized expanding shell found by Martin
(1998); the accumulated dust surrounds HII region 2, where ionized gas 
and dust could be partially mixed within it. 

\begin{figure*}
\begin{minipage}[b]{0.5\textwidth}
\centering
\includegraphics[width=6.4cm]{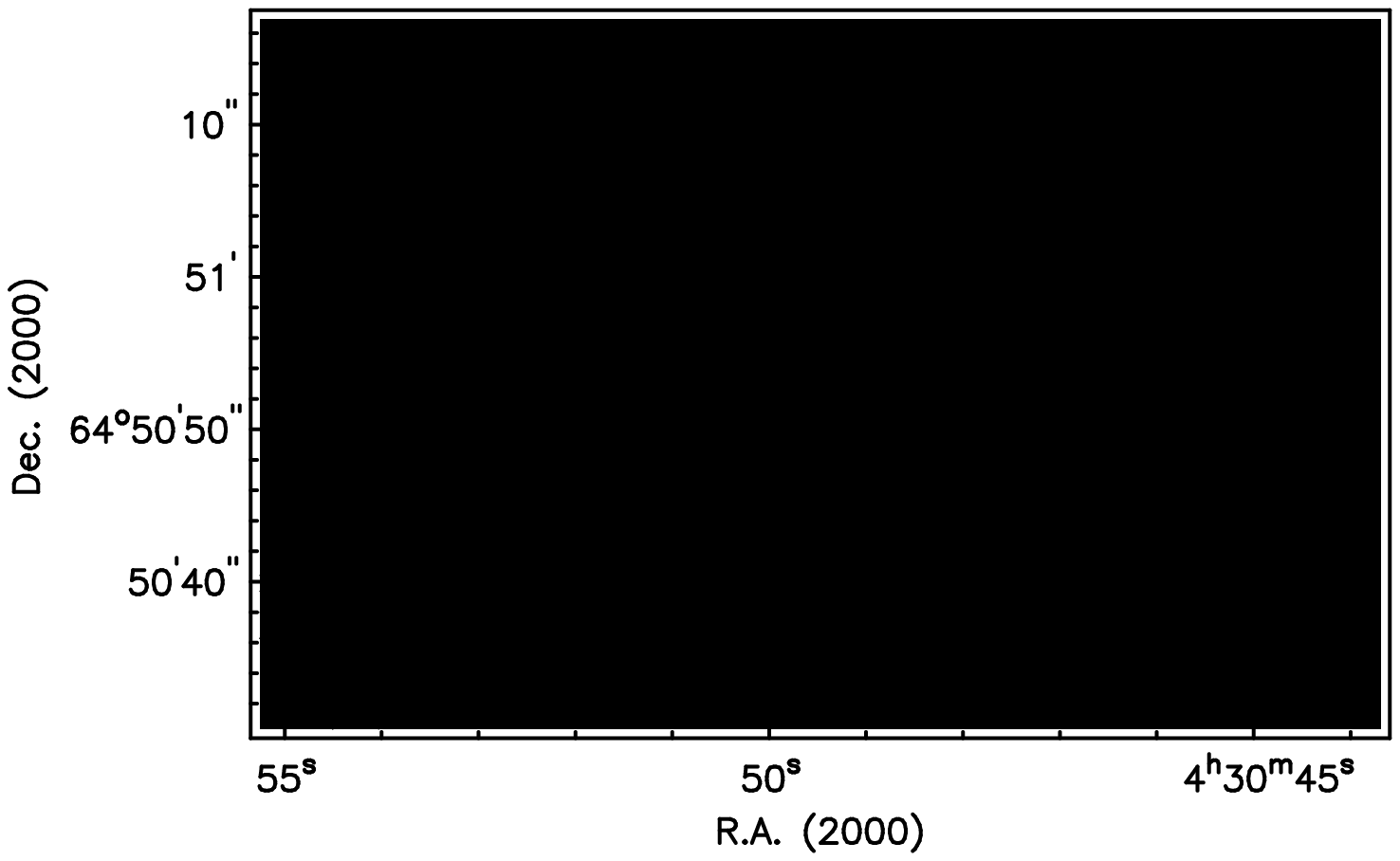}
\end{minipage}
\begin{minipage}[b]{0.5\textwidth}
\centering
\includegraphics[width=6.4cm]{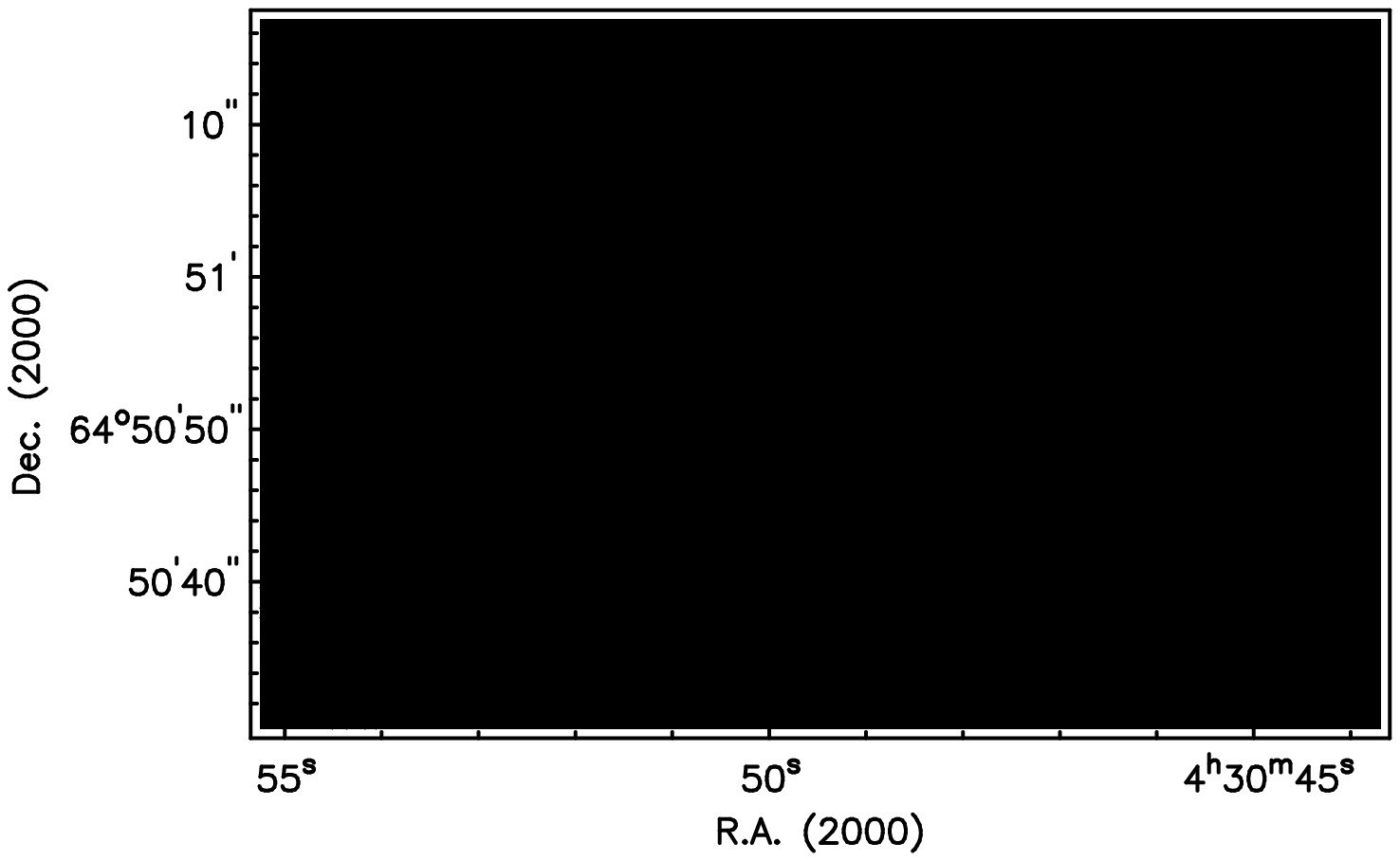}
\end{minipage}
\begin{minipage}[b]{0.5\textwidth}
\centering
\includegraphics[width=6.4cm]{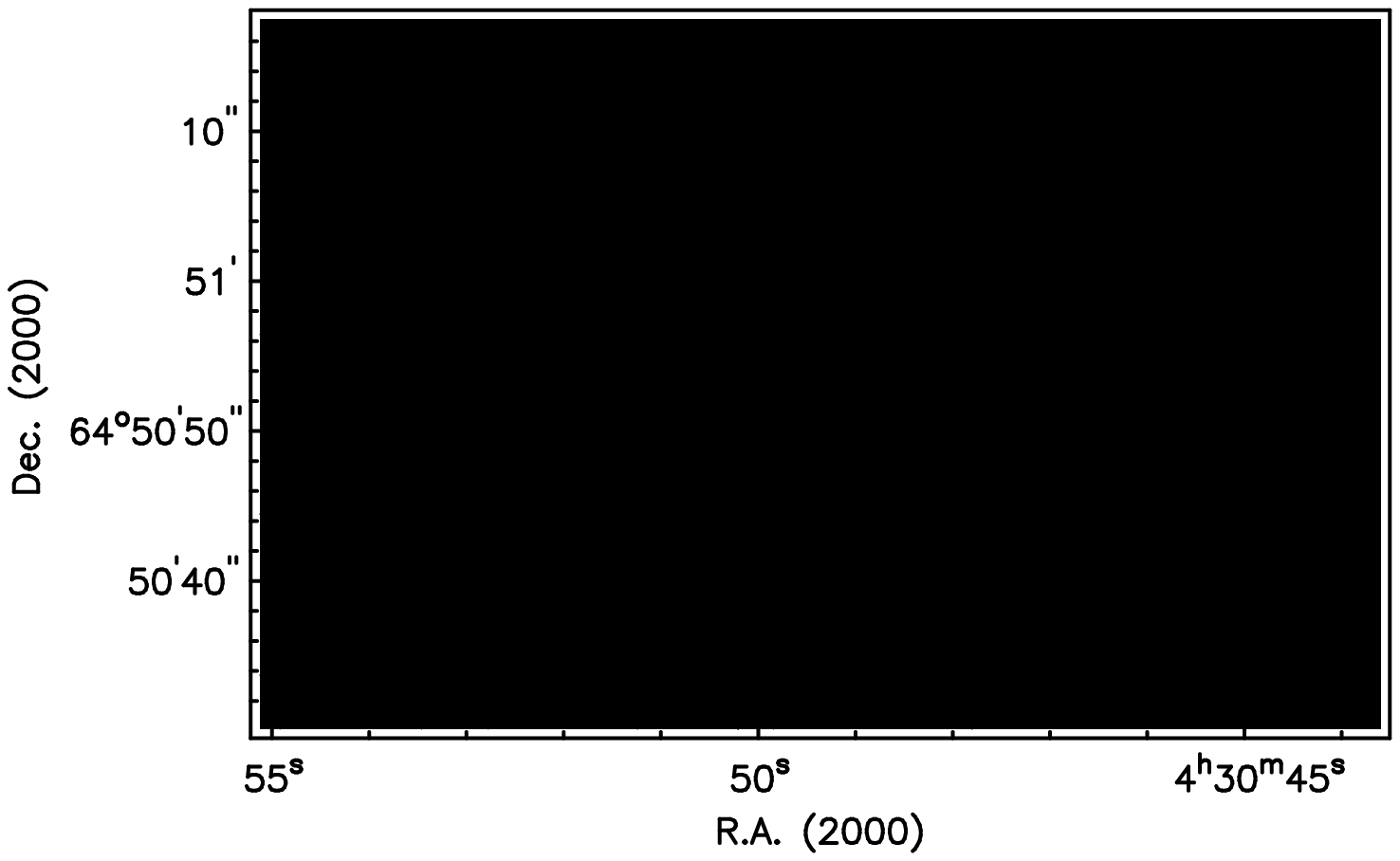}
\end{minipage}
\begin{minipage}[b]{0.5\textwidth}
\centering
\includegraphics[width=6.4cm]{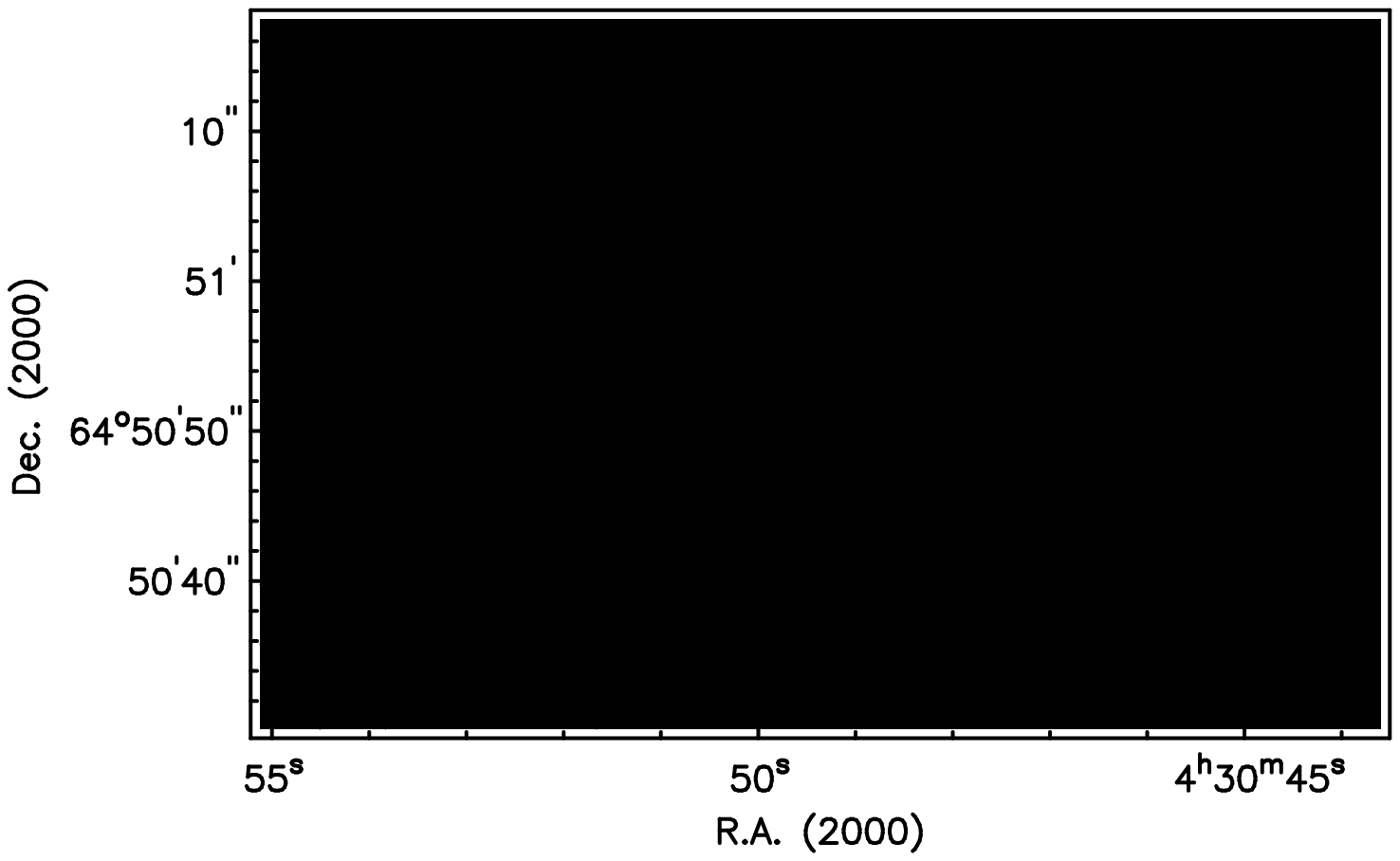}
\end{minipage}
  \caption{{\bf a.} 8$\mu$m (IRAC) intensity map (1.2'') with H$\alpha$
   (2'') intensity overlaid, {\bf b.} the same as a. but with extinction
   contours overlaid (2''). {\bf c.} 24$\mu$m (MIPS) intensity map (2.6'') with H$\alpha$
   (6'') intensity overlaid, {\bf d.} the same as c. but with extinction
   contours overlaid (6'').}
              \label{spitzer_data}
\end{figure*}

In order to further study the distribution of the dust within the ionized
gas, we compared the Balmer extinction and the H$\alpha$
emission at 2'' and 6'' resolution with the dust emission
at 8$\mu$m and 24$\mu$m obtained with the IRAC and MIPS instruments on
SPITZER (see Fig.~\ref{spitzer_data}). We
note the following: {\bf i.} While the peak at 8$\mu$m coincides with the 
maxima in H$\alpha$, there is a displacement between this peak and the
extinction maximum (see Figs.~\ref{spitzer_data}a
and~\ref{spitzer_data}b). A good spatial agreement between H$\alpha$,
24$\mu$m and 8$\mu$m has also been found by Calzetti et al. (2005).  
{\bf ii.} A displacement is also seen at 24$\mu$m, where the maximum is offset
from the extinction maximum and closer to the
H$\alpha$ peak. There might be a slight displacement between the 24$\mu$m and
H$\alpha$ maxima (see Fig.~\ref{spitzer_data}c) but the low
resolution of the 24$\mu$m image does not allow us to make a firm
statement about this. {\bf iii.} Many details of the 8$\mu$m emission
have a perfect correspondence in the extinction map (see Fig.~\ref{spitzer_data}b); e.g. local maxima
at HII regions 5 and 7 and tails 1 and 2 (marked by arrows in
Fig.~\ref{spitzer_data}b). The latter tail (which does not appear in
the 24$\mu$m image although it is clearly seen in the 8$\mu$m image
smoothed at the 24$\mu$m resolution not shown here) coincides
with an HI ridge within the galaxy (see Fig.~11 in Rela\~no et al.
2006). {\bf iv.} Tail 3 in Fig.~\ref{spitzer_data}b, located at the
north, coincides with the western ridge and maximum in HI (Fig.~11 in
Rela\~no et al. 2006), but it does not show up in the extinction
map. This is most likely due to the predominance of cold neutral gas,
as compared to ionized gas, so that the Balmer decrement becomes a poor
tracer for the dust. 

\acknowledgements 
This work has been supported by the Spanish Ministry of Education, via
PNAYA (Spanish National Program for Astronomy and Astrophysics),
the research projects AYA 2005-07516-C02-01, AYA 2004-08260-C03-C02 and ESP 2004-06870-C02-02, and
the Junta de Andaluc\'\i a.

\end{document}